\PassOptionsToPackage{bookmarks=false}{hyperref}
\documentclass[sigconf]{acmart}
\usepackage{float}
\usepackage{enumitem}
\usepackage{graphicx}
\usepackage{subfiles}
\usepackage{array}
\usepackage{amsmath}
\usepackage{algorithm}
\usepackage{algpseudocode}
\usepackage{multirow}
\usepackage{subfigure}
\usepackage{caption}
\usepackage{diagbox}
\usepackage{ulem}
\usepackage{flushend}
\usepackage{balance}
\usepackage{color}
\usepackage{xcolor,colortbl}
\usepackage{float}

\newcommand{\tool}{BugListener}
\newcommand{\numProject}{six}

\copyrightyear{2022} 
\acmYear{2022} 
\setcopyright{acmcopyright}\acmConference[ICSE '22]{44th International Conference on Software Engineering}{May 21--29, 2022}{Pittsburgh, PA, USA}
\acmBooktitle{44th International Conference on Software Engineering (ICSE '22), May 21--29, 2022, Pittsburgh, PA, USA}
\acmPrice{15.00}
\acmDOI{10.1145/3510003.3510108}
\acmISBN{978-1-4503-9221-1/22/05}

\begin{document}

\title{
BugListener: Identifying and Synthesizing Bug Reports from Collaborative Live Chats
}




\author{Lin Shi$^{1,2}$,  Fangwen Mu$^{1,2}$, Yumin Zhang$^{1,2}$,  Ye Yang$^{5}$,  Junjie Chen$^{6}$, Xiao Chen$^{1,2}$, Hanzhi Jiang$^{1,2}$,Ziyou Jiang$^{1,2}$,  Qing Wang$^{1,2,3,4}$}

\affiliation{ $^1$Laboratory for Internet Software Technologies, Institute of Software Chinese Academy of Sciences
\city{Beijing}
\country{China}}
\affiliation{$^2$ University of Chinese Academy of Sciences
\city{Beijing}
\country{China}}
\affiliation{$^3$ State Key Laboratory of Computer Science, Institute of Software Chinese Academy of Sciences
\city{Beijing}
\country{China}}
\affiliation{$^4$ Science \& Technology on Integrated Information System Laboratory, \\ Institute of Software Chinese Academy of Sciences
\city{Beijing}
\country{China}}
\affiliation{$^5$ School of Systems and Enterprises, Stevens Institute of Technology
\city{Hoboken, NJ}
\country{USA}}
\affiliation{$^6$ Tianjin University, College of Intelligence and Computing
\city{Tianjin}
\country{China}
\{shilin,fangwen2020,yumin2020,chenxiao2021,hanzhi2021,ziyou2019,wq\}@iscas.ac.cn,\\
yyang4@stevens.edu,
junjiechen@tju.edu.cn\\
}

\authornote{Corresponding author.\\ }

\renewcommand{\shortauthors}{Lin Shi et al.}

\begin{abstract}

In community-based software development, developers frequently rely on live-chatting to discuss emergent bugs/errors they encounter in daily development tasks. 
However, it remains a challenging task to accurately record such knowledge due to the noisy nature of interleaved dialogs in live chat data.
In this paper, we first formulate the task of identifying and synthesizing bug reports from community live chats, and propose a novel approach, named {\tool}, to address the challenges.
Specifically, {\tool} automates three sub-tasks: 1) Disentangle the dialogs from massive chat logs by using a Feed-Forward neural network; 
2) Identify the bug-report dialogs from separated dialogs by 
leveraging the Graph neural network to learn the contextual information;
3) Synthesize the bug reports by utilizing {Transfer Learning techniques} to classify the sentences into: observed behaviors (OB), expected behaviors (EB), and steps to reproduce the bug (SR).
{\tool} is evaluated on six open source projects. The results show that: for bug report identification, {\tool} achieves the average F1 of 77.74\%, improving the best baseline by 12.96\%; and for bug report synthesis task, {\tool} could classify the OB, EB, and SR sentences with the F1 of {84.62\%, 71.46\%, and 73.13\%}, improving the best baselines by {9.32\%, 12.21\%, 10.91\%}, respectively.
A human evaluation study also confirms the effectiveness of {\tool} in generating relevant and accurate bug reports.
These demonstrate the significant potential of applying {\tool} in community-based software development, for promoting bug discovery and quality improvement.

\end{abstract}


\maketitle

\section{Introduction}
Collaborative communication via live chats 
allows developers to seek information and technical support, share opinions and ideas, discuss issues, and form community development \cite{chatterjee2019exploratory, DBLP:journals/jss/ChatterjeeKP20},
in a more efficient way compared with asynchronous communication such as emails or forums \cite{DBLP:conf/cscw/LinZSS16,DBLP:conf/msr/ShihabJH09,DBLP:conf/icsm/ShihabJH09}. Consequently, collaborative live chatting has become an integral part of most software development processes, not only for open source communities 
constituting globally distributed developers, but also for software companies to facilitate in-house team communication and coordination, esp. in accommodating remote work due to the COVID-19 pandemic \cite{DBLP:journals/corr/abs-2101-05877}. 

\begin{figure}[t]
\centering
\includegraphics[width=0.9\columnwidth,height=8.6cm]{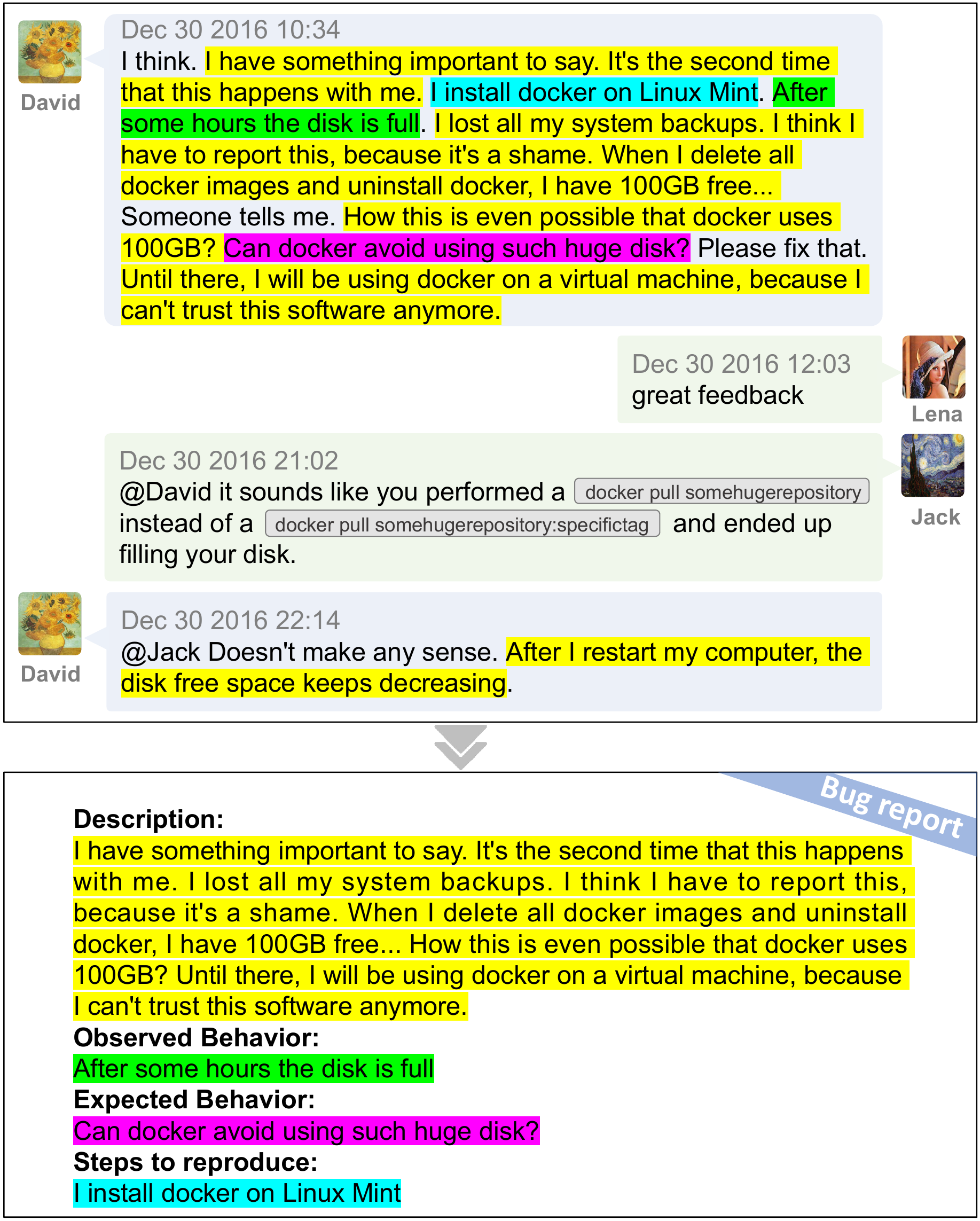}
\vspace{-0.2cm}
\caption{An example of identifying and synthesizing a bug report from the Docker collaborative live chats.}
\label{fig:motivation}
\vspace{-0.5cm}
\end{figure}

Existing literature reports that developers are likely to join collaborative live chats to discuss problems they encountered during development \cite{DBLP:conf/msr/ParraEH20,DBLP:conf/msr/ChatterjeeDKP20,DBLP:conf/msr/AlkadhiLGB17,DBLP:conf/wcre/AlkadhiNGB18}. Shi et al. \cite{shi2021first} analyzed 749 
{live-chat dialogs} from eight OSS communities, and found 32\% of the dialogs are reporting unexpected behaviors, such as something does not work, reliability issues, performance issues, and errors. In fact, these reporting problems usually imply potential bugs that have not been found. Fig. \ref{fig:motivation} illustrates an example slice of collaborative live chats \cite{docker} from the Docker community. 
In this conversation, developer \textit{David} reported a performance bug that Docker took a lot of disk space, and Lena indeed confirmed \textit{David}'s feedback.
Then, \textit{Jack} provided a suggestion to help resolve this problem but failed in the end.
Although developers have revealed this bug via collaborative live chats, the {highly dynamic and multi-threading}
nature of live chatting makes this bug-report conversation get quickly flooded by new incoming messages.
After several months, Docker developers call to remembrance this bug with the 
frustrated comments such as 
``lost all my system backups'' and ``it's a shame'',
when there are several formal bug reports (i.e., \href{https://github.com/moby/moby/issues/30254}{\#30254}, \href{https://github.com/moby/moby/issues/31105}{\#31105}, and \href{https://github.com/moby/moby/issues/32420}{\#32420}) reflecting the similar problem that was submitted to the GitHub bug repository.
We can observe that, if the bug discussed in live chats could be identified and documented in a timely manner, the bug may have been resolved earlier by the Docker community. Consequently, the Docker community may have the opportunity to {prevent many failure incidents associated with this bug} 
\cite{DBLP:journals/infsof/PitangueiraTSMB17}. 


Although the live chats could be a tremendous data source embedded with bug reports over time, it is quite challenging to mine massive chat messages due to the following barriers. 
\textbf{(1) Entangled and noisy data}. Live chats 
typically contain entangled, informal conversations covering a wide range of topics \cite{DBLP:conf/ijcai/LiuSGLWZ20}.
Moreover, there exist noisy utterances such as duplicate and off-topic messages in chat messages that do not provide any valuable information. 
Such entangled and noisy nature of live chat data poses a difficulty in analyzing and interpreting the communicative dialogues. 
\textbf{(2) Understanding complex dialog structure}.
In complex dialogs, developers usually either confirm or reject a bug report by replying to previous utterances. Since the ``reply-to'' relationship is not linear to the dialog structure, it is necessary to employ more sophisticated techniques to handle nonlinear dialog structure, in order to learn precise feedback and reduce the likelihood of introducing false-positive. 
For example, the utterance ``When I use the `automationName' key, I get an error that it is not a recognized W3C capability.'' is very likely to be classified as a bug proposal.
However, when examining the dialog, we found that the following-up utterances pointed out the error was not a valid bug. Instead, it was caused by the user's action of importing incorrect packages.
\textbf{(3) Extremely expensive annotation}. The live chats are typically large in size. It is extremely expensive to annotate bug reports from chat messages due to the high volume corpus and a low proportion of ground-truth data. Only a few labeled chat messages are categorized into bug report types. 
 Thus, the labeled resources for synthesizing bug reports are also limited. How to make maximal use of the limited labeled data to classify the unlabeled chat messages accurately becomes a critical problem.

In this work, we propose a novel approach, named {\tool}, which can identify bug-report dialogs from massive chat logs and synthesize complete bug reports from predicted bug-report dialogs. 
{\tool} employs a deep graph-based network to capture the complex dialog structure, and a transfer-learning network to synthesize bug reports.
Specifically, {\tool} addresses the challenges with three elaborated sub-tasks:
1) Disentangle the dialogs from massive chat logs by using a Feed-Forward neural network.
2) Identify bug-report dialogs from separated dialogs by modeling the original dialog to the graph-structured dialog and leveraging the Graph neural network (GNN) to learn the complex context representation.
3) Synthesize the bug reports from predicted bug-report dialogs using Transfer Learning techniques. {Specifically, we use the pre-trained BERT model provided by Devlin et al. \cite{DBLP:conf/naacl/DevlinCLT19} and fine-tune it twice using the external BEE dataset \cite{song2020bee} and our own dataset, respectively.}
To evaluate the proposed approach, we collect and annotate 1,501 dialogs from six popular open-source projects. The experimental results show that our approach significantly outperforms all other baselines in both two tasks. For bug report identification task, {\tool} achieves an average F1 of 77.74\%, improving the best baseline by 12.96\%. For bug report synthesis task, {\tool} could classify sentences depicting observed behavior (OB), expected behavior (EB), and steps to reproduce (SR) with the F1 of {84.62\%, 71.46\%, and 73.13\%}, respectively, improving the best baseline by {9.32\%, 12.21\%, and 10.91\%}, respectively.
We also conduct a human evaluation to assess the correctness and quality of the generated bug reports, showing that {\tool} can generate relevant and accurate bug reports.

The main contributions and their significance are as follows. 
\begin{itemize}[leftmargin=*]
    \item We propose an automated approach, named {\tool}, based on a deep graph-based network to effectively identify the bug-report dialogs, and a transfer-learning network to extensively synthesize bug reports. We believe that {\tool} can \textbf{facilitate community-based software development by promoting bug discovery and quality improvement}.
    \item We evaluate the {\tool} by comparing with state-of-the-art baselines, with superior performance.
    \item \textbf{Data availability}: publicly accessible dataset and source code \cite{website} to facilitate the replication of our study and its application in other contexts. 
\end{itemize} 

In the remaining of this paper, 
Sec. 2 defines the problem.
Sec. 3 elaborates the approach. 
Sec. 4 presents the experimental setup. Sec. 5 demonstrates the results and analysis. 
Sec. 6 describes the human evaluation.
Sec. 7 discusses indications and threats to validity. 
Sec. 8 introduces the related work. Sec. 9 concludes our work.
\section{Problem Definition}
\label{sec:background}

To facilitate the problem definition and further discussion, we first provide some basic concepts and notations used in this study:

\begin{itemize}[leftmargin=*]
    \item \textbf{A chat log (L)} corresponds to a sequence of utterances $u_i$ in chronological order, denoted by $L=\{u_1,u_2,...,u_n\}$.
    \item \textbf{An utterance ($u_i$)} consists of the timestamp, developer role, and textual message, denoted by $u_i = < time, role, text>$.
    \item \textbf{A developer role ($role$)} in a dialog is defined as either a \textit{reporter} or a \textit{discussant}. A \textit{reporter} refers to a developer launching a dialog, while a \textit{discussant} refers to a developer participating in the dialog. Denoted by $ role \in \{reporter, discussant\}$.
    \item \textbf{A dialog ($D_i$)} is a sequence of $k$ utterances $u_i$, retaining the ``reply-to'' relationship among utterances,  
    denoted by $D_i = \{u_1^{\mathcal{R}_1},\\ u_2^{\mathcal{R}_2},...,u_k^{\mathcal{R}_k}\}$. 
    \item \textbf{A relational context for utterance $u_i$ ($\mathcal{R}_i)$} 
    is a set of {undirected "reply-to"} relationship identifiers, each identifier corresponding to a message {replying} to {or replied by} $u_i$. {If two utterances {share} the same superscript, then {it implies one replies to the other}.}
    For example, $D=\{u_1^{R1,R2},u_2^{R1},u_3^{R2}\}$ represents that both $u_2$ and $u_3$ reply to $u_1$.  
\end{itemize}

Our work then targets at automatically identifying and synthesizing bug reports from community live chats. We formulate the task of automatic bug report generation from live chats with three elaborated sub-tasks:

(1) Dialog disentanglement: Given the historical chat log $L$, disentangle it into separate dialogs $\{D_1,D_2,...,D_n\}$.

(2) Bug-Report dialog Identification (BRI): Given a separate dialog $D_i$, find a binary function $f$ so that $f(D_i)$ can determine whether the dialog {involves bug-reporting messages}.

(3) Bug-Report Synthesis (BRS):
Assuming that the content of bug reports is made up of sentences extracted from the reporters' utterances, 
given all the reporter's utterances $U_r$ in the predicted bug-report dialog $D_i$, 
find a function $g$ so that $g(U_r)=\{DES,OB,EB,SR\}$, 
where $DES, OB, EB,$ and $SR$ represent the collections of 
sentences in $U_r$ that depict \textit{Description, Observed Behavior, Expected Behavior}, and \textit{Step to Reproduce}.

\section{Approach} 
\begin{figure*}[t]
\centering
\vspace{-0.3cm}
\includegraphics[width=\textwidth]{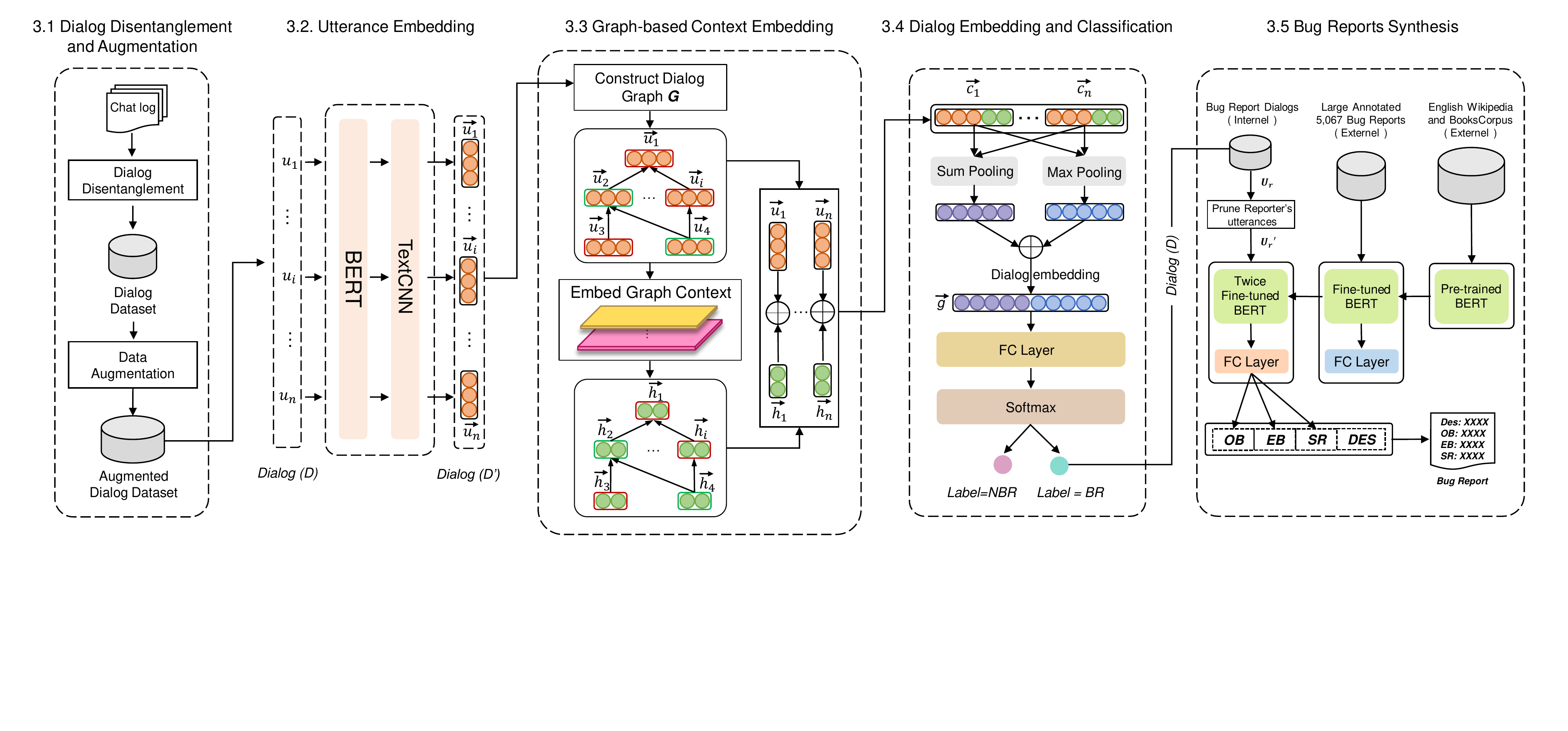}
\vspace{-2.8cm}
\caption{Overview of {\tool}.}
\label{fig:approach}
\vspace{-0.4cm}
\end{figure*}

There are five main steps to construct {\tool}, as shown in Fig. \ref{fig:approach}. These include:(1) dialog disentanglement and data augmentation to prepare the data; (2) utterance embedding to convert utterances into semantic vectors; (3) graph-based context embedding to construct dialog graph and learn the contextual representation by employing a two-layer graph neural network; (4) dialog embedding and classification to learn whether a dialog is a bug-report dialog; and (5) bug report synthesis to form a complete bug report. Next, we present details of each step.

\vspace{-0.15cm}

\subsection{Data Disentanglement and Augmentation}

In this step, We first separate dialogs from the interleaved chat logs using a Feed-Forward network. Then, we augment the original dialog dataset utilizing a heuristic data augmentation method to overcome the insufficient labeled resource challenge. 
\vspace{-0.1cm}
\subsubsection{Dialog Disentanglement}
Utterances from a single conversation thread are usually interleaved with other ongoing conversations, and therefore need to be divided into individual dialogs accordingly.
{To find a reliable disentanglement model, we experiment with four state-of-the-art dialog disentanglement models, i.e., BILSTM model \cite{BILSTMMODEL}, BERT model \cite{DBLP:conf/naacl/DevlinCLT19}, E2E model \cite{DBLP:conf/ijcai/LiuSGLWZ20}, and FF model, using our manual disentanglement dataset {as detailed in Section 4.1 later}. The comparison results from our experiments show that the FF model significantly outperforms the others on disentangling developer live chat by achieving the highest scores on NMI, Shen-F, F1, and ARI metrics. The average scores of these four metrics are 0.74, 0.81, 0.47, and 0.57  respectively\footnote{Due to space, experimental details on evaluation existing disentanglement models are provided on our website \cite{website}.}.}


Specifically, the FF model is a Feed-Forward neural network with 2 layers, 512-dimensional hidden vectors, and softsign non-linearities. 
It employs a two-stage strategy to resolve dialog disentanglement.
First, the FF model predicts the ``reply-to'' relationship between every two utterances in the chat log based on averaged pre-trained word embedding and many hand-engineered features. 
Second, it clusters the utterances that can reach each other via the ``reply-to'' predictions as one dialog. {Thus, the FF-model can output not only the utterances in one dialog but also their ``reply-to'' relationship, which is essential for constructing {the internal network structure of} dialogs.}

\subsubsection{Data Augmentation}
To address the limited annotation and data imbalance issue, a heuristic data augmentation mechanism is employed to enlarge the dataset through dialog mutation. 
The key to dialog mutation is to alter the utterance forms and retain their semantics.
To achieve that, we mutate a long utterance by replacing a few words with their synonyms, or mutate a short utterance by replacing it with another short utterance. Specifically,
given a dialog $D=\{u_1,u_2,...,u_n\}$, we generate $N$ different mutants by iterating the following steps $N$ times. 
For each utterance $u_i$ in a dialog $D$, we perform either an utterance-level replacement or a word-level replacement based on its length, and generate a new utterance ${u_i}'=\Gamma(u_i)$:
\begin{equation}
\forall u_i \in D, \Gamma(u_i)=
\begin{cases}
u_k& |u_i| \leq \theta\\
\textit{SR}(u_i)& |u_i| > \theta
\end{cases}
\end{equation}
where $|u_i|$ denotes the length of $u_i$, and $\theta$ is a predefined threshold (We empirically set $\theta=5$ in this study). $u_k$ is the utterance that is randomly selected from the entire dialog corpus with a length less than $\theta$.
$\textit{SR}(u_i)$ denotes the synonym-replacement operation that has been widely used by NLP text augmentation task \cite{DBLP:conf/emnlp/WeiZ19}.
After all utterances in dialog $D$ are processed, we then obtain a new dialog 
 $D_{aug}=\{{u_1}',{u_2}',...,{u_n}'\}$.

To achieve data balancing, for each project, we first augment the NBR dialogs to a certain number, then we augment BR dialogs to {match} the same number.
Taking the Angular project as an example,
we first augment the NBR dialogs from 179 to 358 (2 times), then augment the  BR dialogs from 86 to 358 for balancing purposes.

\subsection{Utterance Embedding}
The utterance embedding aims to encode semantic information of words, as well as to learn the representation of utterances.

\textbf{Word encoding.} 
We encode each word in utterances into a semantic vector by utilizing the deep pre-trained BERT model \cite{DBLP:conf/naacl/DevlinCLT19}, which has achieved impressive success in many natural language processing tasks \cite{task1, task2}. The last layer of the BERT model outputs a 768-dimensional contextualized word embedding for each word.

\textbf{Utterance encoding.} 
With all the word vectors, we use TextCNN \cite{TextCNN} to learn the utterance representation. TextCNN is a classical method for sentence encoding by using a shallow Convolution Neural Network (CNN) \cite{CNN} to learn sentence representation. It has an advantage over learning on insufficient labeled data, since it employs a concise network structure and a small number of parameters.
We use four different size convolution kernels with 100 feature maps in each kernel. The convoluted features are fed to a \textit{Max-Pooling} layer followed by the \textit{ReLU} activation \cite{Relu}. Then, we concatenate these features and input them into a 100-dimensional full-connected layer to obtain the 100-dimensional utterance embedding $\vec{u_i}$.
After encoding all the utterances of a dialog $D$,  we can get utterance-embedded dialog $D'=\{\vec{u_1},\vec{u_2},..,\vec{u_n}\}$.



\vspace{-0.1cm}
\subsection{Graph-based Context Embedding}

This step aims to capture the graphical context of utterances in one dialog. 
Given the utterance-embedded dialog $D'=\{\vec{u_1},\vec{u_2},..,\vec{u_n}\}$ with the set of ``reply-to'' relationship $\mathcal{R}$, we first construct a dialog graph $G(D')$. Then, we learn the contextual information of $G(D')$ via a two-layer graph neural network, and output $G_c(D')$ where each vertex in $G_c(D')$ restores the contextual information of the corresponding vertex in $G(D')$. Finally, We concatenate each vertex in $G(D')$ with its corresponding vertex in $G_c(D')$, and output the sequence of combination as the dialog vector $C = \{\vec{c_1},\vec{c_2},...,\vec{c_n}\}$.

\vspace{-0.1cm}
\subsubsection{Construct Dialog Graph}

Given the utterance-embedded dialog $D'$ consisting of $N$  utterances and the set of ``reply-to'' relationship $\mathcal{R}$, we construct a directed graph $G(D') = (\mathcal{V},\mathcal{E},\mathcal{W},\mathcal{T})$, 
where $\mathcal{V}$ is the vertex set, $\mathcal{E}$ is the edge set, $\mathcal{W}$ is the weight set of edges, and $\mathcal{T}$ is the set of edge types. More specifically:

\textbf{Vertex.} 
Each utterance is represented as a vertex $v_i \in \mathcal{V}$. We use the utterance embedding $\vec{u_i}$ to initialize the corresponding vertex $v_i$. The $v_i$  will be updated during the graph learning process.

\textbf{Edge.} 
We construct the edge set $\mathcal{E}$ based on the ``reply-to'' relationship. The edge $e_{ij} \in \mathcal{E}$ denotes that there is a ``reply-to'' relationship between $u_i$ and $u_j$.

\textbf{Edge Weight.} The edge weight $w_{ij}$ 
is the weight of the edge $e_{ij}$, with $0 \leq w_{ij} \leq 1$, where $w_{ij} \in \mathcal{W}$ and $i,j\in[1,2,...,N]$. 
$w_{ij}$ is determined by the similarity of $\vec{u_i}$ and $\vec{u_j}$. Specifically, we employ pair-wise dot product to compute the similarity score of pair vertices. Then, we normalize the similarity score and calculate the edge weight $w_{ij}$:
\begin{equation}
    w_{ij}= \frac{\vec{u_i}^\mathrm{\emph{T}}\cdot W_e\vec{u_j}}{\sum\limits_{k \in N_{(i,*)}}{\vec{u_i}^\mathrm{\emph{T}}\cdot W_e\vec{u_k}}}
\end{equation}
where $W_e$ is a trainable matrix used to perform linear feature transformation on vertex, $N_{(i,*)}$ denotes the set of vertices that vertex $v_i$ points to.

\textbf{Edge Type.} 
We define the type of the edge $e_{ij}$ as $t_{ij} \in \mathcal{T}$, according to the \textit{developer-role dependency} of $e_{ij}$. Specifically, we consider four types of edges in this study, i.e.,  
$r \rightarrow r, r \rightarrow d, d \rightarrow r$, and $d \rightarrow d$, 
where $r$ denotes the reporter, $d$ denotes the discussant, as we defined in {the previous section.}

\subsubsection{Embed Dialog Graph Context}
Given a dialog graph $G(D')$, we employ a two-layer graph neural network (GNN) \cite{GNN} to embed the graph context of dialog structure and developer-role dependency, respectively. We output $G_c(D')$ where each vertex restores graph context information.

\textbf{Structure-level GNN.} 
In the first layer, a basic GNN \cite{basicGNN} is used to learn the structure-level context for each vertex in a given graph, including embedding its neighbor vertices via the ``reply-to'' edges, as well as the features contained in the neighbor vertices.

A basic GNN layer can be implemented as follows:
\begin{equation}
    {v_i}^{(l+1)}=\sigma\bigg(W_1^{(l)}{v_i}^{(l)} + W_2^{(l)}\sum\limits_{j \in N_{(*,i)}}{v_j}^{(l)}\bigg)
\end{equation}
where $N_{(*,i)}$ denotes the set of neighboring vertices that point to vertex $v_i$. ${v_i}^{(l)}$ represents the {updated}
vertex at layer $l$, and ${v_i}^{(l+1)}$ represents the updated vertex at layer $l+1$. $\sigma$ denotes a non-linear function, such as {sigmoid} or {ReLU}, $W_1^{(l)}$ and $W_2^{(l)}$ are trainable parameter matrices. We introduce the edge weights to better aggregate the local information. Hence, the updated vertex ${v_i}^{(1)}$ of the {structure-level GNN} layer is calculated as:
\begin{equation}
    {v_i}^{(1)}=\sigma\bigg(W_1^{(1)}\vec{u_i} + W_2^{(1)}\sum\limits_{j \in N_{(*,i)}}w_{ji}\vec{u_j}\bigg)
\end{equation}
where $w_{ji}$ denotes the edge weight from vertex $v_j$ to vertex $v_i$.



\textbf{Role-level RGCN.}
In the second layer, we further capture the high-level contextual information by leveraging Relational Graph Convolutional Networks (RGCN) \cite{RGCN}. 
RGCN is a generalization of Graph Convolutional Networks (GCN) \cite{GCN} which extends the hierarchical propagation rules and takes the edge types between vertices into account. Since RGCN explicitly models the neighborhood structures, it can better handle multi-relational graph data like our dialog graph, which contains four edge types. The  vertex  ${v_i}$ is updated by applying the RGCN over the output of the first layer.
\begin{equation}
    {v_i}=\sigma\bigg(W_1^{(2)}{v_i}^{(1)} + \sum\limits_{t \in T}\sum\limits_{j \in N_{(*,i)}^{t}}\frac{1}{c_{i,t}}W_t^{(2)}{v_j}^{(1)}\bigg)
\end{equation}
where $N_{(*,i)}^{t}$ denotes the set of vertices that point to vertex $v_i$ under edge type $t \in \mathcal{T}$. $c_{i,t}$ is a normalization constant that can either be learned or set in advance (such as $c_{i,t} = \vert N_{(*,i)}^{t}\vert$). $\sigma$ denotes a non-linear function, $W_2^{(l)}$ and $W_t^{(l)}$ are trainable parameter matrices, the latter matrix changes under different edge types.
{The output of role-level RGCN is the $G_c(D')$ where each vertex $v_i$  restores the embedded graph context $\vec{h_i}$ for utterance $u_i$.} 

\subsubsection{Combined representation.} 

To enrich the utterance representation,
we concatenate each vertex in $G(D')$ with its corresponding vertex in $G_c(D')$, and output the sequence of combination as the dialog vector $C = \{\vec{c_1},\vec{c_2},...,\vec{c_n}\}$, where $\vec{c_i} = [\vec{u_i}\oplus\vec{h_i}]$, and $\oplus$ is the concatenation operator.

\subsection{Dialog Embedding and Classification}
This step aims to obtain the representation of an entire dialog and classify it as either a positive or a negative bug-report dialog.

\textbf{Dialog Embedding.} We input the dialog vector $C = \{\vec{c_1},\vec{c_2},...,\vec{c_n}\}$ to the \textit{Sum-Pooling} and the \textit{Max-Pooling} layer respectively. 
Then, we concatenate the output vectors to get the dialog embedding $\vec{g}$:
\begin{equation}
    \vec{g}=\sum\limits_{i=1}^{\vert V\vert}\vec{c_i}\oplus\text{Maxpooling}(\vec{c_1},...,\vec{c_n})
\end{equation}
where $\oplus$ is the concatenation operator, $\vert V\vert$ is the number of the graph's vertices.

\textbf{Dialog Classification.} The label is predicted by feeding the dialog embedding $\vec{g}$ into two \textit{Full-Connected} (FC) layers followed by the \textit{Softmax} function:
\begin{equation}
   \mathcal{P}=\text{softmax}(FC_2(\text{ReLU}(FC_1(\vec{g_e}))))
\end{equation}
where $\mathcal{P}$ is the 2-length vector [$P(\text{NBR}\vert D),P(\text{BR}\vert D)$], the $P(\text{NBR}\vert D)$ is the predicted probability of non-bug-report dialog, the $P(\text{BR}\vert D)$ is the predicted probability of bug-report dialog.

Finally, we minimize the loss through the Focal Loss \cite{focalloss} function. The Focal Loss improves the standard Cross-Entropy Loss by adding a focusing parameter $\gamma \geq 0$. It focuses on training on hard examples, while down-weight the easy examples.
\begin{equation}
    FL = -\sum\limits_{i} \alpha_i(1-\mathcal{P}_i)^\gamma y_i\log(\mathcal{P}_i)
\end{equation}
where $y_i$ is the $i$-th element of the one-hot ground-truth label (BR or NBR), $\alpha_i$ and $\gamma$ are tunable parameters.

\subsection{Bug Report Synthesis}

Due to the high volume of live chat data and the low proportion of ground-truth bug-report dialogs, it is difficult to get enough training data for bug report synthesis task. {To address this challenge, we utilize a twice fine-tuned BERT model, which proves to be effective to improve performance through {more sophisticated} 
transferring knowledge from the pre-trained model \cite{DBLP:conf/naacl/DevlinCLT19}. Specifically, we use a pre-trained BERT and fine-tune it twice using the external BEE dataset and our BRS dataset, as shown in the dashed box of `3.5' in Fig. \ref{fig:approach}.}




\textbf{(1) Initial Fine-tuning BERT model.}
{The BERT model is a bidirectional transformer using a combination of \textit{Masked Language Model} and \textit{Next Sentence Prediction}. It is trained from English Wikipedia (2,500M words) and BooksCropus (800M words) \cite{Zhu_2015_ICCV}. The entire BERT model is a stack of 12 BERT layers with more than 100 million parameters.}

{Based on an assumption that the contents of bug reports are likely from the reporters' utterances, we perform the initial fine-tune on the task of classifying bug-report contents into OB, EB, SR, and Others.} 
First, we select the external BEE dataset proposed by Song et al. \cite{song2020bee} that includes 5,067 bug reports, 11,776 OB sentences, 1,568 EB sentences, and 24,655 SR sentences as the source dataset. Second, following the previous study \cite{song2020bee}, we preprocess sentences in the 5,076 bug reports with lowercase, tokenization, excluding non-English and overlong (over 200 words) ones.
Third, {we freeze the first nine layers of the pre-trained BERT and update the parameters of the last three layers via the sentences in the 5,076 bug reports. We take the output of the first token (the [\emph{CLS}] token) as the sentence embedding.}
Finally, we input the sentence embedding into a FC layer  to produce the probabilities of OB ($P_b$), EB ($P_e$), SR ($P_s$), and Others ($P_o$). 
We apply \textit{Cross-Entropy} Loss when measuring the difference between truth and prediction:
\begin{equation}
    Loss = -(y_{b}log(P_{b})+y_{e}log(P_{e})+y_{s}log(P_{s}) + y_{o}log(P_{o}))
\end{equation}
where $y_{b}$, $y_{e}$, $y_{s}$, and $y_{o}$ indicate the ground-truth labels of sentences.

\textbf{(2) 
Twice fine-tuning BERT model.}
{Given the above fine-tuned BERT model, we perform the second round of fine-tuning on our BRS dataset as follows.} We first collect all the reporter's utterances $U_r$ in Dialog $D$ as our inputs.   
Since $U_r$ may contain trivial contents that are less meaningful for reporting bugs, we prune the $U_r$ into $U_r'$ if they satisfy the following heuristic rules: 
(1) remove the sentence $s$ if: ($length(s) \leq 5$) AND ($s$ does not contain [URL], [EMAIL], [HTML], [CODE] or [VERSION]) (2) remove the string $str$ from its sentence if: $\forall str \in $ \{``Hi'', ``Hi All'', ``hey there'', ``Hi everybody'', ``hey  guys'', ``hi guys'', ``guys'', ``Hi there'', ``thank\ you'', ``thanks'', ``thanks\ anyway'', ``thanks\ for\ replaying'', ``ok, thanks'', {etc.}\}. 
Second, we transfer the BERT model previously fine-tuned on the external bug report dataset for initialization, and replace the original FC layer with a new one. 
{Third, 
the {BERT} model is fine-tuned the second time via labeled sentences in $U_r'$ using a smaller learning rate}. 

\textbf{(3) Bug reports assembling.} When generating bug reports, we assemble sentences that are predicted to the same category in chronological order. To fully retain the useful information in $U_r'$, we assemble all the sentences that belong to the ``Others'' category as the description paragraph. In the end, we could generate a bug report with its description, observed behavior, expected behavior, and step to reproduce according to best practices for bug reporting
\cite{zimmermann2010makes, DBLP:conf/sigsoft/BettenburgJSWPZ08}.

\section{Experimental Design}
To evaluate the proposed {\tool} approach, our evaluation specifically addresses three research questions:

\textbf{RQ1}: How effective is {\tool} in identifying bug-report dialogs from live chat data? 

\textbf{RQ2}: How effective is {\tool} in synthesizing bug reports? 

\textbf{RQ3}: How does each individual component in {\tool} contribute to the overall performance? 

\begin{table*}[htbp]
\caption{Our Experiment Dataset. (Part, Dial, Uttr, Sen are short for participating developers, dialog, utterance, and sentence, respectively. BR and NBR denote bug-report and non-bug-report dialogs. $U_r$ denotes  sentences in reporter's utterances, and $U_r'$ denotes the pruned $U_r$.)}
\vspace{-0.3cm}
\label{dataset}
\centering
\resizebox{\textwidth}{!}{%
\begin{tabular}{|c|c|c|c|c|c|c|c|p{1cm}<{\centering}|c|p{1cm}<{\centering}|c|c|c|p{0.7cm}<{\centering}|c|c|c|c|c|}
\hline
                                         & \multicolumn{3}{c|}{}                                             & \multicolumn{4}{c|}{\textbf{Sample Population}}                      & \multicolumn{4}{c|}{\textbf{BRI Dataset}}                                                & \multicolumn{8}{c|}{\textbf{BRS Dataset}}                                                                                                      \\ \cline{5-20} 
\multirow{-2}{*}{\textbf{}}              & \multicolumn{3}{c|}{\multirow{-2}{*}{\textbf{Entire Population}}} & \multicolumn{2}{c|}{\textbf{NBR}} & \multicolumn{2}{c|}{\textbf{BR}} & \multicolumn{2}{c|}{\textbf{Augmented NBR}} & \multicolumn{2}{c|}{\textbf{Augmented BR}} & \multicolumn{2}{c|}{\textbf{BR Dialog}} & \multicolumn{2}{c|}{\textbf{Reporter Sen.}} & \multicolumn{4}{c|}{\textbf{BR content}}               \\ \hline
\cellcolor[HTML]{FFFFFF}\textbf{Project} & \textbf{Part.}        & \textbf{Dial.}       & \textbf{Uttr}       & \textbf{Dial.}  & \textbf{Uttr.}  & \textbf{Dial.}  & \textbf{Uttr}  & \textbf{Dial.}       & \textbf{Uttr.}       & \textbf{Dial.}       & \textbf{Uttr}       & \textbf{Dial.}      & \textbf{Sen.}     & \textbf{Ur}          & \textbf{Ur'}         & \textbf{OB} & \textbf{EB} & \textbf{SR} & \textbf{DES} \\ \hline
Angular                                  & 22,467                & 79,619              & 695,183             & 179              & 1,043             & 86              & 268            & 358                  & 2,086                & 358                  & 1,132               & 86                  & 647               & 507                  & 446                  & 177          & 34           & 40          & 195          \\ \hline
Appium                                   & 3,979                 & 4,906               & 29,039              & 169              & 737             & 84              & 233            & 338                 & 1,474               & 338                  & 935             & 84                  & 596               & 478                  & 397                  & 180          & 29           & 44           & 144           \\ \hline
Docker                                   & 8,810                 & 3,964               & 22,367              & 172              & 916             & 61              & 185             & 344                  & 1,832                & 344                 & 1,037               & 61                  & 438               & 367                  & 322                  & 150          & 35           & 32          & 105           \\ \hline
DL4J                                     & 8,310                 & 27,256              & 252,846             & 178              & 1,070             & 79              & 373            & 356                  & 2,140                & 356                  & 1,781               & 79                  & 828               & 590                  & 502                  & 184          & 32           & 56           & 230          \\ \hline
Gitter                                   & 9,260                 & 7,452               & 34,147              & 207              & 813             & 63              & 304             & 414                 & 1,626                & 414                  & 1,898               & 63                  & 733               & 432                  & 369                   & 159          & 19           & 15          & 176           \\ \hline
Typescript                               & 8,318                 & 18,812              & 196,513             & 203              & 1,016             & 20              & 85             & 406                  & 2,032                & 406                  & 1,625               & 20                  & 176              & 138                   & 118                   & 48          & 12           & 8           & 50           \\ \hline
\textit{\textbf{Total}}                  & 61,144                & 142,009             & 1,230,095           & 1,108             & 5,595           & 393             & 1,448            & 2,216                & 11,190               & 2,216                & 8,408              & 393                 & 3418             & 2,512                  & 2,154                  & 898         & 161          & 195          & 900          \\ \hline
\end{tabular}
\vspace{-0.3cm}
}
\end{table*}

\subsection{Data Preparation}

\subsubsection{Studied Communities}

Many OSS communities utilize Gitter \cite{gitter.im} or Slack \cite{slack.com} as their live communication means. 
Considering the popular, open, and free access nature, we select studied communities from {Gitter}\footnote{In Slack, communities are controlled by the team administrators, whereas in Gitter, access to the chat data is public}.
{Following previous work \cite{DBLP:conf/kbse/PanBR00L21, DBLP:conf/kbse/ShiJYCZMJW21} , we select popular and active communities as our studied subjects.
Specifically,} we select the Top-1 most participated communities from {\numProject} active domains, covering front end framework, mobile, data science, DevOps, collaboration, and programming language.
Then, we collect the live chat utterances from these communities. Gitter provides REST API \cite{gitter.rest} to get data about chatting rooms and post utterances. In this study, we use the REST API to acquire the chat utterances of the {\numProject} selected communities, and the retrieved dataset contains all utterances as of ``2020-12-31''. 

\subsubsection{Preprocessing and Disentanglement}
For data preprocessing, we first convert all the words in utterances into lowercase, and remove the stopwords. We also normalize the contractions in utterances with contractions \cite{contractions} library and use Spacy \cite{spacy.io} for lemmatization. Following previous work \cite{boutet2021emojis, suman2021emoji}, we replace the emojis with specific strings to standard ASCII strings. Besides, we detect low frequency tokens such as URL, email address, code, HTML tag, and version number with regular expressions, and substitute them into [URL], [EMAIL], [HTML], [CODE], and [VERSION] respectively. Then, we use the FF model \cite{disentangle2} to divide the processed data into individual dialogs as introduced in Sec. 3.1.1. 
The detailed statistic is shown in the ``Entire Population'' column of Table \ref{dataset}.

\subsubsection{Sampling and Filtering}
After dialog disentanglement, the number of individual chat dialogs remains quite large. Limited by the human resource of labeling, we randomly sample 100 dialogs from each community. 
{The sample population accounts for about 1.1\% of the entire population. Although the ratio is not large, we consider the selected dialogs are representative because they are randomly selected from six diverse communities.}
The details of sampling results are shown in the ``Sample Population'' column of Table \ref{dataset}.

{Since {\tool} relies on natural language processing to understand the dialog, dialogs that have too much noise or do not contain enough information are almost incomprehensible and thus cannot decide a bug report.} {Following the data cleaning procedures of previous studies \cite{DBLP:conf/kbse/PanBR00L21, DBLP:conf/kbse/ShiJYCZMJW21},} we excluded noisy dialogs {by applying the following exclusion criteria}: 1) Dialogs that are written in non-English languages; {2) Dialogs where the code or stack traces accounts for more than 90\% of the entire chat content;} 3) Low-quality dialogs such as dialogs with many typos and grammatical errors. 4) Dialogs that involve channel robots {which main handle simple greeting or general information messages}.

\subsubsection{Ground-truth Labeling}
{For each sampled dialog {obtained in the previous step},} we label ground-truth data from three aspects: 
(1) Correct disentanglement results. For each sampled dialog, we manually correct the prediction of the ``reply-to'' relationships between utterances, as well as the disentanglement results.
(2) Label dialogs with BR and NBR (See the ``Sample Population'' column in Table 1). For each dialog that has been manually corrected, we manually label it with a ``BR'' or an ``NBR'' tag, according to whether it discusses a certain bug that should be reported.
(3) Label sentences with OB, EB, and SR (See the ``BRS Dataset'' column in Table 1). For each dialog labeled with BR, we first prune all reporter's utterances $U_r$ to obtain $U_r'$ as described in Sec. 3.5(2). Then we label each sentence in $U_r'$ with observed behavior (OB), expected behavior (EB), and step to reproduce (SR), according to their contents.

To ensure the labeling validity, we built an inspection team, which consisted of four PhD students. All of them are fluent English speakers, and have done either intensive research work with software development or have been actively contributing to open-source projects. We divided them into two groups. The results from both groups were cross-checked and reviewed.
When a labeled result received different opinions, we hosted a discussion with all team members to decide through voting. 
Based on our observation, the correctness of automated dialog disentanglement is 79\%. 
The average Cohen’s Kappa about bug report identification is 0.87, and the average Cohen’s Kappa about bug report synthesis is 0.84.

\subsubsection{Dataset augmentation and balancing}
For BRI task, we augment the dataset as introduced in Sec. 3.1.
For each project, we first augment the NBR data eight times, and then augment the BR data until BR and NBR data are balanced. The details are shown in the ``BRI Dataset'' column in Table 1.
For BRS task, we apply {EDA\cite{DBLP:conf/emnlp/WeiZ19} techniques to augment} OB, EB, SR sentences until their numbers are balanced.
We further incorporate an external dataset for transfer learning. The external dataset is provided by Song et al. \cite{song2020bee}, including 5,067 bug reports with 11,776 OB sentences, 1,568 EB sentences, and 24,655 SR sentences.

\begin{table*}[t]
\caption{Baseline comparison across the six communities for bug-report dialog identification (\%). }
\vspace{-0.2cm}
\centering
\label{tab:rq1}
\resizebox{\textwidth}{!}{%
\begin{tabular}{|c|c|c|c|c|c|c|c|c|c|c|c|c|c|c|c|c|c|c|c|c|c|}
\hline
                                   & \multicolumn{3}{c|}{\textbf{Angular}}                                                      & \multicolumn{3}{c|}{\textbf{Appium}}                                                       & \multicolumn{3}{c|}{\textbf{Docker}}                                                          & \multicolumn{3}{c|}{\textbf{DL4J}}                                                           & \multicolumn{3}{c|}{\textbf{Gitter}}                                                       & \multicolumn{3}{c|}{\textbf{Typescript}}                                                      & \multicolumn{3}{c|}{\textbf{Average}}                                                         \\ \cline{2-22} 
\multirow{-2}{*}{\textbf{Methods}} & P                             & R                             & F1                         & P                             & R                          & F1                            & P                             & R                             & F1                            & P                             & R                            & F1                            & P                             & R                          & F1                            & P                             & R                             & F1                            & P                             & R                             & F1                            \\ \hline
\textbf{{\tool}}                      & \cellcolor[HTML]{FFEB9C}82.93 & \cellcolor[HTML]{FFEB9C}79.07 & \cellcolor[HTML]{FFEB9C}80.95 & 69.39                         & 80.95 & \cellcolor[HTML]{FFEB9C}74.73 & \cellcolor[HTML]{FFEB9C}77.42 & \cellcolor[HTML]{FFEB9C}78.69                         & \cellcolor[HTML]{FFEB9C}78.05 & \cellcolor[HTML]{FFEB9C}85.07                         & \cellcolor[HTML]{FFEB9C}72.15 & \cellcolor[HTML]{FFEB9C}78.08                         & \cellcolor[HTML]{FFEB9C}82.09                         & \cellcolor[HTML]{FFEB9C}87.30 & \cellcolor[HTML]{FFEB9C}84.62 & \cellcolor[HTML]{FFEB9C}70.00                         & \cellcolor[HTML]{FFEB9C}70.00 & \cellcolor[HTML]{FFEB9C}70.00 & \cellcolor[HTML]{38FFF8}77.82 & \cellcolor[HTML]{38FFF8}78.03 & \cellcolor[HTML]{38FFF8}77.74 \\ \hline
\textbf{NB}                        & 58.88                         & 73.26                         & 65.28                      & 62.22                         & 66.67                         & 64.37                         & 65.52                            & 31.15 & 42.22                         & 62.79                         & 34.18 & 44.26                         & 72.92                         & 55.56                      & 63.06                         & 35.29                         & 30.00                         & 32.43                         & 59.60                         & 48.47                        & 51.94                         \\ \hline
\textbf{GBDT}                      & 72.22                         & 60.47                         & 65.82                         & 65.17 & 69.05                         & 67.05                         & 66.00                         & 54.10                         & 59.46                         & 85.00                         & 64.56 & 73.38                         & 59.77 & 82.54                      & 69.33                         & 35.14                            & 65.00                         & 45.61                         & 63.88                         & 65.95                         & 63.44                         \\ \hline
\textbf{RF}                        & 75.00                          & 59.30                         & 66.23                      & \cellcolor[HTML]{FFEB9C}72.15                         & 67.86                         & 69.94                         & 68.75                         & 36.07                         & 47.31                         & 72.73                         & 20.25                         & 31.68                            & 62.34                         & 76.19                         & 68.57                         & 60.00 & 30.00                         & 40.00                         & 68.50                         & 48.28                         & 53.96                         \\ \hline
\textbf{FastText}                  & 77.59                         & 52.33                         & 62.50                      & 68.54                         & 72.62                         & 70.52                         & 56.60                          & 49.18                         & 52.63                         & 74.51                         & 48.10                        & 58.46                            & 67.24                         & 61.90 & 64.46                         & 40.91                         & 45.00 & 42.86                         & 64.23                        & 54.86                         & 58.57                         \\ \hline
\textbf{CNC}                       & 80.36                         & 52.33                         & 63.38                      & 67.05                            & 70.24                         & 68.60                            & 74.51                         & 62.29                        & 67.86                         & 84.44 & 48.10 & 61.29 & 68.18                         & 71.43                      & 69.77                         & 52.00                         & 65.00 & 57.78                          & 71.09                         & 61.57                         & 64.78                         \\ \hline
\textbf{DECA}                      & 51.32                        & 45.35                         & 48.15                      & 51.16                         & 52.38                         & 51.76                         & 45.57                            & 59.02                            &51.43                            & 42.22                  & 48.10                            & 44.97                         & 55.36                      & 49.20                         & 52.10                         & 21.57                         & 55.00                          & 30.99                         & 44.53        & 51.51                 & 46.57                         \\ \hline
\textbf{Casper}                    & 67.65                            & 53.49                         & 59.74                      & 66.06                         & \cellcolor[HTML]{FFEB9C}85.71                         & 74.61                         & 60.56                         & 70.49                         & 65.15                         & 82.14                         & 58.23                        & 68.15                         & 73.33                          & 69.84                      & 71.54                         & 32.50                            & 65.00                         & 43.33                         & 63.71                        & 67.13                         & 63.75                         \\ \hline
\end{tabular}%
}
\vspace{-0.3cm}
\end{table*}

\subsection{Baselines}
The first two RQs require comparison with state-of-the-art baselines.  
We employ four common machine-learning-based baselines applicable to both RQ1 and RQ2, {including {\textbf{Naive Bayesian (NB)}} \cite{mccallum1998comparison}, {\textbf{Random Forest (RF)}} \cite{liaw2002classification}, {\textbf{Gradient Boosting Decision Tree (GBDT)}} \cite{ke2017lightgbm}, and \textbf{FastText} \cite{joulin2016fasttext}.} 
In addition, we employ several baselines applicable to RQ1 and RQ2, respectively.

{\textbf{Additional Baselines for identifying bug-report dialogs ( RQ1).}} 
{Furthermore, we also consider some existing approaches that can identify sentences or mini-stories which are discussing problems.}
\textbf{CNC \cite{Huang2018Automating}}
is the state-of-the-art learning technique to classify sentences in comments taken from online issue reports. They proposed a CNN \cite{CNN}-based approach to classify sentences into seven categories of intentions: Feature Request, Solution Proposal, Problem Discovery, etc. To achieve better performance of the CNC baseline, we retrain the CNC model on our BRI dataset. 
We assemble all the utterances in a dialog as an entry, and predict whether the entry belongs to problem discovery.
\textbf{DECA} \cite{DBLP:conf/kbse/SorboPVPCG15}
is the state-of-the-art rule-based technique 
for analyzing development emails. It is used to classify the sentences of emails
into problem discovery, solution proposal, information giving, etc., by using linguistic rules. We use the twenty-eight linguistic rules \cite{deca_web} for identifying the ``problem discovery'' utterances in a dialog and regard the dialog containing the ``problem discovery'' utterances as the bug-report dialog.
\textbf{Casper \cite{DBLP:conf/icse/0002S20}}
is a method for extracting and synthesizing user-reported mini-stories regarding app problems from reviews. Similar to the CNC baseline, we also retrain the Casper model on the BRI dataset, and apply it to determine bug-report dialogs by assembling all the utterances in a dialog as one entry.

{\textbf{Additional Baseline for synthesizing bug reports (RQ2).}}
We investigated seven state-of-the-art approaches for the bug report synthesis task, including CUEZILLA \cite{zimmermann2010makes}, DeMlBUD \cite{chaparro2017detecting}, iTAPE \cite{DBLP:conf/kbse/ChenXYJCX20}, S2RMiner \cite{DBLP:conf/icsr/ZhaoMYZP19}, infoZilla \cite{bettenburg2008extracting},
Euler \cite{DBLP:conf/sigsoft/ChaparroBLMMPPN19}
and BEE \cite{song2020bee}. 
Among the above approaches, only the replication packages from iTAPE, S2RMiner, and BEE are available. Since iTAPE and S2RMiner classify SR sentences, and only BEE share the same target with us, that is to classify \textit{OB, EB, SR,} and \textit{Other} sentences for bug reports.
Therefore, we choose BEE as our additional baselines for bug report synthesis. BEE comprises three binary classification SVM, which can tag sentences with OB, EB, or SR labels.

{This leads to a total of seven baselines for RQ1, and five baselines for RQ2.}

\subsection{Evaluation Metrics}
We use three commonly-used metrics to evaluate the performance of both two tasks, i.e., \textit{Precision, Recall, and F1}.
(1) \textit{Precision} refers to the ratio of the number of correct predictions to the total number of predictions; 
(2) \textit{Recall} refers to the ratio of the number of correct predictions to the total number of samples in the golden test set; and (3) \textit{F1} is the harmonic mean of precision and recall. When comparing the performances, we care more about F1 since it is balanced for evaluation.

\subsection{Experiment Settings}
The experimental environment is a desktop computer equipped with an NVIDIA GeForce RTX 3060 GPU, intel core i5 CPU, 12GB RAM, running on Ubuntu OS.

For RQ1, we apply \textit{Cross-Project Evaluation} on our BRI dataset to perform the training process. We iteratively select one project as a test dataset, and the remaining five projects for training. We train {\tool} with 32 \textit{batch\_size}. We choose \textit{Adam} as the optimizer with \textit{learning\_rate}=1e-4. To avoid over-fitting, we set \textit{dropout}=0.5, and adopt the \textit{L2-regularization} with $\lambda$=1e-5. The $\alpha$ and $\gamma$ of \textit{Focal Loss} function are 0 and 2, respectively. 
When training GBDT, we set the \textit{learning\_rate}=0.1 and the \textit{n\_estimators}=100;
For RF, we set the \textit{min\_samples\_leaf}=10 and the \textit{n\_estimators}=100;
We train 100 epochs for FastText, and set the \textit{learning\_rate}=0.1, the window size of input n-gram as 2;
Casper chooses \textit{SVM.SVC} as the default function, with \textit{rbf} as the kernel, 3 as the degree, and 200 as the \textit{cache\_size}; CNC selects 32 as the \textit{batch\_size}, 128-dimensional word embedding, four different filter sizes of $[2,3,4,5]$ with 128 filters, 30 training epochs, and \textit{dropout}=0.5.
For these hyper-parameters, we use greedy search \cite{DBLP:conf/kbse/LiSYW20} as the parameter selection method to obtain the best performance.

For RQ2, {in the first fine-tune round, we train {\tool} on the external BEE dataset (see Sec. 4.1.5)} with {64} \textit{batch\_size}. {We set the warmup proportion of BERT model to 0.1, and the value of gradient clip to 1.0.}
We choose \textit{Adam} as the optimizer with {\textit{learning\_rate}=1e-4 and \textit{weight decay rate}=0.01. We train {\tool} for 13 epochs and save the best model.} 
{In the second fine-tune round,} we use the same parameters while changing the \textit{batch\_size} {from 64 to 8, the \textit{epoch} from 13 to 70, and the \textit{learning\_rate} from 1e-4 to 1e-6. }We apply a \textit{10-fold partition} on the BRS dataset to perform the {secondary} fine-tuning, i.e., we use nine folds for fine-tuning, and the remaining one for testing. For NB/GDBT/RF/FastText baselines, we use the greedy strategy to tune parameters to achieve the best performance. For the additional baseline BEE, we directly utilize its open API \cite{bee} to predict OB, EB, and SR sentences.

For RQ3, we compare {\tool} with its two variants in bug report identification task: 
1) \textbf{{\tool} w/o CNN}, which removes the TextCNN.
2) \textbf{{\tool} w/o GNN}, which removes the graph neural network. {\tool} with its two variants use the same parameters when training.
{We compare {\tool} with its variant without transferring knowledge from the external BEE dataset {(i.e., \textbf{{\tool} w/o TL})} in bug report synthesis task.
{\tool} w/o TL has the same network structure with {\tool}, but it does not use the external BEE dataset and is only fine-tuned on our BRS dataset.}

\section{Results and Analysis}
\label{sec:result}
\subsection{Performance in Identifying Bug Reports}

Table \ref{tab:rq1} shows the comparison results between the performance of {\tool} and those of the seven baselines across data from six OSS communities, for \textbf{BRI} tasks. 
The columns correspond to Precision, Recall, and F1. 
The highlighted cells indicate the best performance from each column.
Then, we conduct the normality test and T-test between every two methods. Overall, the data follows a normal distribution, and {\tool} significantly ($p-value < 0.01$) outperforms the seven baselines on F1.
Specifically, when comparing with the best Precision-performer among the seven baselines, i.e., CNC, {\tool} can improve its average precision by 6.73\%. Similarly, {\tool} improves the best Recall-performer, i.e., Casper, by 10.90\% for average recall, and improves the best F1-performer, i.e., CNC, by 12.96\% for average F1. At the individual project level, {\tool} can achieve the best F1-score in all six communities. 

{For \textbf{BRI tasks}, }we believe that the performance advantage of {\tool} is mainly attributed to the rich representativeness of its internal construction, from two perspectives:
(1) {\tool} models the textual dialog as the dialog graph thereby can effectively exploit the graph-structured knowledge. While the structure information is {missing} in the baseline methods that {treat} a dialog as a {linear structure}.
(2) {\tool} leverages a novel two-layer GNN model with considering the edge types between utterances to learn a high-level contextual representation. Thus it can capture the latent semantic relations between utterances more accurately.

\textbf{Answering RQ1:} 
On average, {\tool} has the best precision, recall, and F1, i.e., 77.82\%, 78.03\%, and 77.74\%, improving the best F1-baseline CNC by 12.96\%. On individual projects, it also outperforms the other baselines with achieving the best F1-score in all six communities.

\subsection{Performance in Synthesizing Bug Reports}

\begin{figure}[t]
\centering
\includegraphics[width=\columnwidth]{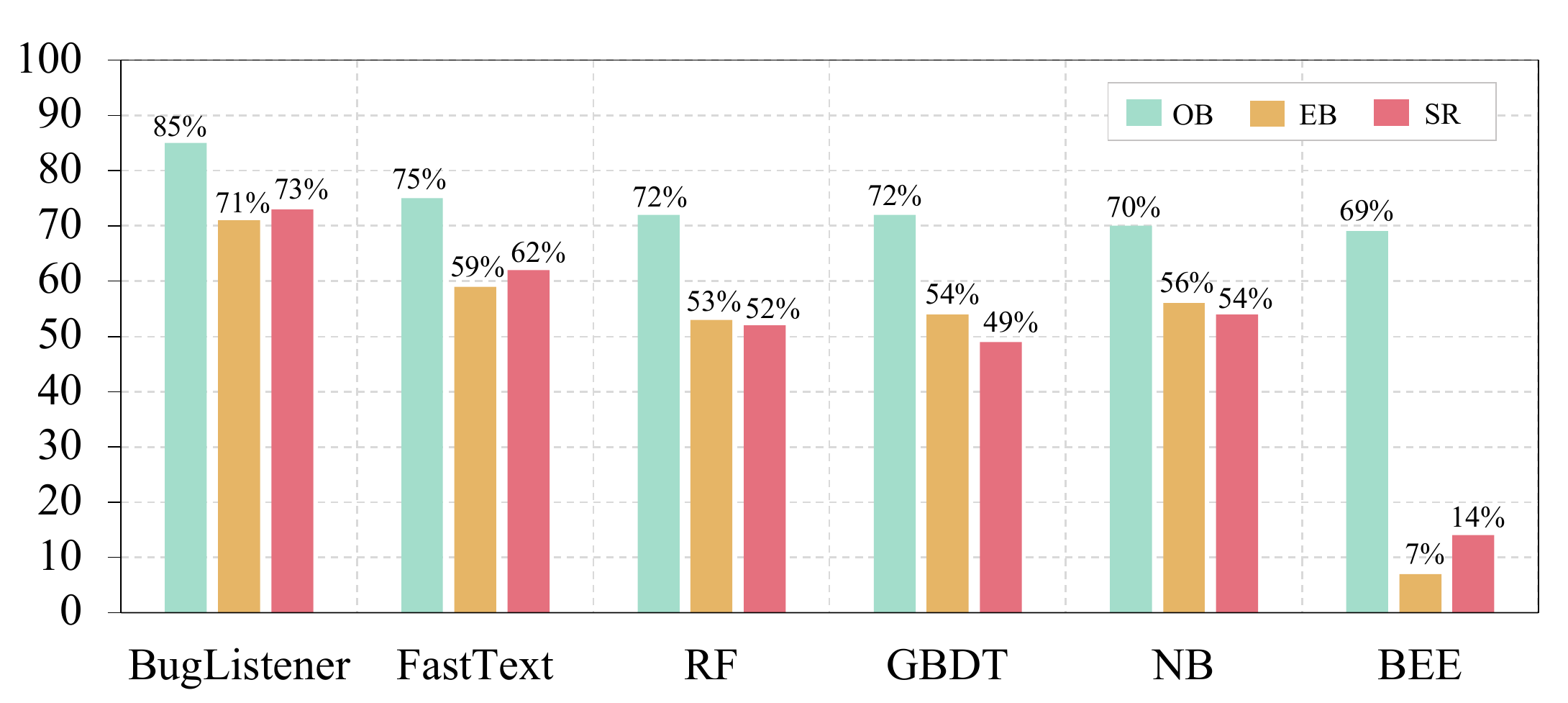}
\vspace{-0.3cm}
\caption{Baseline comparison for bug report synthesis.}
\label{fig:rq2}
\vspace{-0.4cm}
\end{figure}

Fig. \ref{fig:rq2} summarizes the comparison results between the average performance of {\tool} and the five baselines, for \textbf{BRS} task.
We can see that, {\tool} can achieve the highest performance in predicting OB, EB, and SR sentences.
It outperforms the six baselines in terms of F1. For predicting OB sentences, it reaches the highest F1 ({84.63\%}), improving the best baseline {FastText} by {9.32\%}. For predicting EB sentences, it reaches the highest F1 ({71.46\%}), improving the best baseline FastText by {12.21\%}. For predicting SR sentences, it reaches the highest F1 (73.13\%), improving the best baseline FastText by {10.91\%}. 

Our approach is more effective to classify OB, EB, and SR sentences in live chats than others, mainly due to two reasons:
(1) By leveraging the transfer learning technique, {\tool} can obtain general knowledge from existing bug reports, thus would further boost the classification performances on the limited resource.
(2) {By employing the state-of-the-art BERT model which 
has a strong ability to learn semantics via the transformer structure}, {\tool} can capture richer semantic features in word and sentence vectors. 

We notice that FastText achieve the second performances.
These results are mainly due to that, 
FastText can better understand the context by capturing the neighbor words using a fixed-size window when embedding words. 
we also notice that BEE performs the {worst} on predicting EB (average F1 is only {7\%}).
These results are mainly due to that, 
BEE is trained from the external normal bug reports dataset, and
the expression style for EB sentences is quite different between those in normal bug reports and those in live conversations. The EB sentences in bug reports are likely expressed in a declarative tone that state the reporter's expectation as an objective fact, e.g., ``I wish docker can save disk usage''. While in live chats, EB sentences are more likely expressed in an interrogative tone that the reporters inquiry or ask for a reply, e.g., ``Can docker avoid using such huge disk?''. Therefore, it is difficult for BEE to predict EB sentences correctly on live chat data.

\textbf{Answering RQ2:} {\tool} outperforms the six baselines in predicting OB, EB, and SR sentences in terms of F1. The three categories' average Precision, Recall, and F1 are {75.57\%, 77.70\%, and 76.40\%}, respectively.

\subsection{Effects of Main Components}

\begin{figure}[t]
\centering
\subfigure[The BRI performance]{
        \label{fig:BRI}
        \includegraphics[width=0.68\columnwidth]{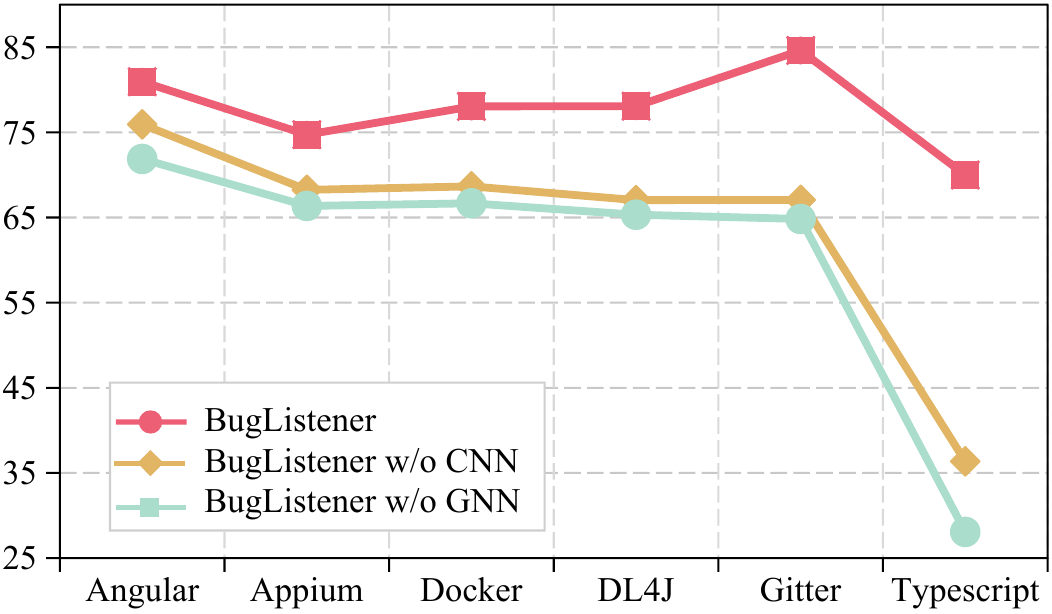}
}
\subfigure[The BRS performance]{
        \label{fig:BRS}
        \includegraphics[width=0.7\columnwidth]{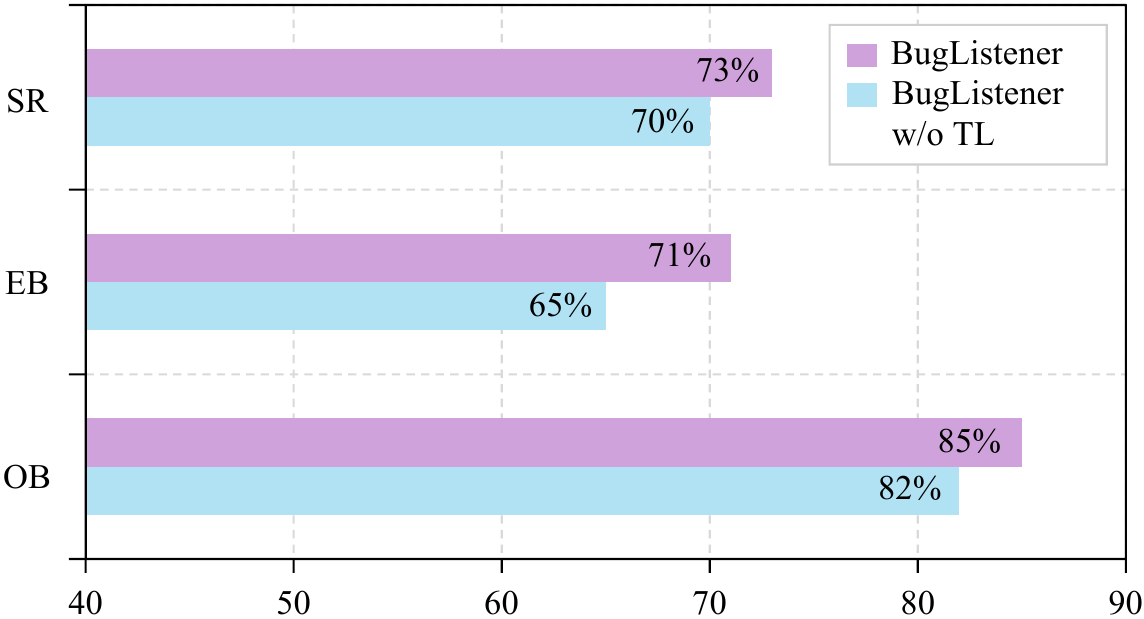}
    
}
\vspace{-0.4cm}
\caption{The component analysis.}
\vspace{-0.4cm}
\label{fig:rq3}
\end{figure}

Fig. \ref{fig:rq3} (a) presents the performances of {\tool} and its two variants for \textbf{BRI} task.
We can see that, the F1 performance of {\tool} is higher than all two variants across all the six communities. When compared with {\tool} and {{\tool} w/o GNN}, removing the GNN component will lead to a dramatic decrease of the average F1 (by 17.22\%) across all the communities. This indicates that the GNN is an essential component to contribute to {\tool}'s high performances. When compared with {\tool} and {{\tool} w/o CNN}, removing the TextCNN component will lead to the average F1 declines by 13.85\%. It is mainly because the TextCNN model can capture the intra-utterance semantic features, which improves the classification performance.

Fig. \ref{fig:rq3} (b) shows the performance of {\tool} and its variant {without transferring knowledge from the external BEE dataset} for \textbf{BRS} task. We can see that, {without the knowledge transferred from the external BEE dataset}, the F1 will averagely decrease by {2.52\%, 6.06\%, 3.13\%} for OB, EB, and SR prediction, respectively.
This indicates that {incorporating the transferred external knowledge} can largely increase the performance on {EB} prediction, while slightly increase the performance on OB and {SR} prediction.

\textbf{Answering RQ3:} The GNN, TextCNN, and Transfer Learning technique adopted by {\tool} are helpful for bug report identification and synthesis.


\vspace{-0.3cm}
\section{Human Evaluation}

To further demonstrate the generalization and usefulness of our approach, we apply {\tool} on recent live chats from five new communities: Webdriverio, Scala, Materialize, Webpack, and Pandas (note that these are different from our studied communities so that all data of these communities do not appear in our training/testing data). Then we invite nine human annotators to assess the correctness, quality, and usefulness of the bug reports generated by {\tool}.

\textbf{Human Annotators}. {We recruit nine participants, including two PhD students, two master students, three professional developers and two senior researchers, all familiar with the five open source communities. They all have at least three years of software development experience, and four of them have more than ten years of development experience.}

\textbf{Procedure}. First, we crawl the recent one-month (July 2021 to August 2021) live chats of the five new communities from Gitter, which contain 3,443 utterances. 
Second, we apply {\tool} to disentangle and construct the live chats into about 562 separated dialogs. Among them, {\tool} identifies \textbf{31 potential bug reports} in total\footnote{Limited by the space, we list the details about the 31 bug reports on our website \cite{website}.}. For each participant, we assign 9-11 bug reports of the communities that they are familiar with. Each bug report is evaluated by three participants. For each bug report, each participant has the following information available: (1) the associated open source community; (2) the original textual dialogs from Gitter; (3) the bug report generated by {\tool}. 

The survey contains three questions:
(1) \textbf{Correctness}: Whether the dialog is discussing a bug that should be reported at that moment (Yes or No)?   
(2) \textbf{Quality}: How would you rate the quality of Description, Observed Behavior, Expected Behavior, and Step to Reproduce in the bug report (using a five-level Likert scale \cite{chisnall1993questionnaire})?
(3) \textbf{Usefulness}: How would you rate the usefulness of {\tool} (using a 5-level Likert scale)?

\begin{figure}[b]
\vspace{-0.6cm}
\centering
\subfigure[Correctness]{
        \label{fig:Correctness}
        \includegraphics[width=0.65\columnwidth]{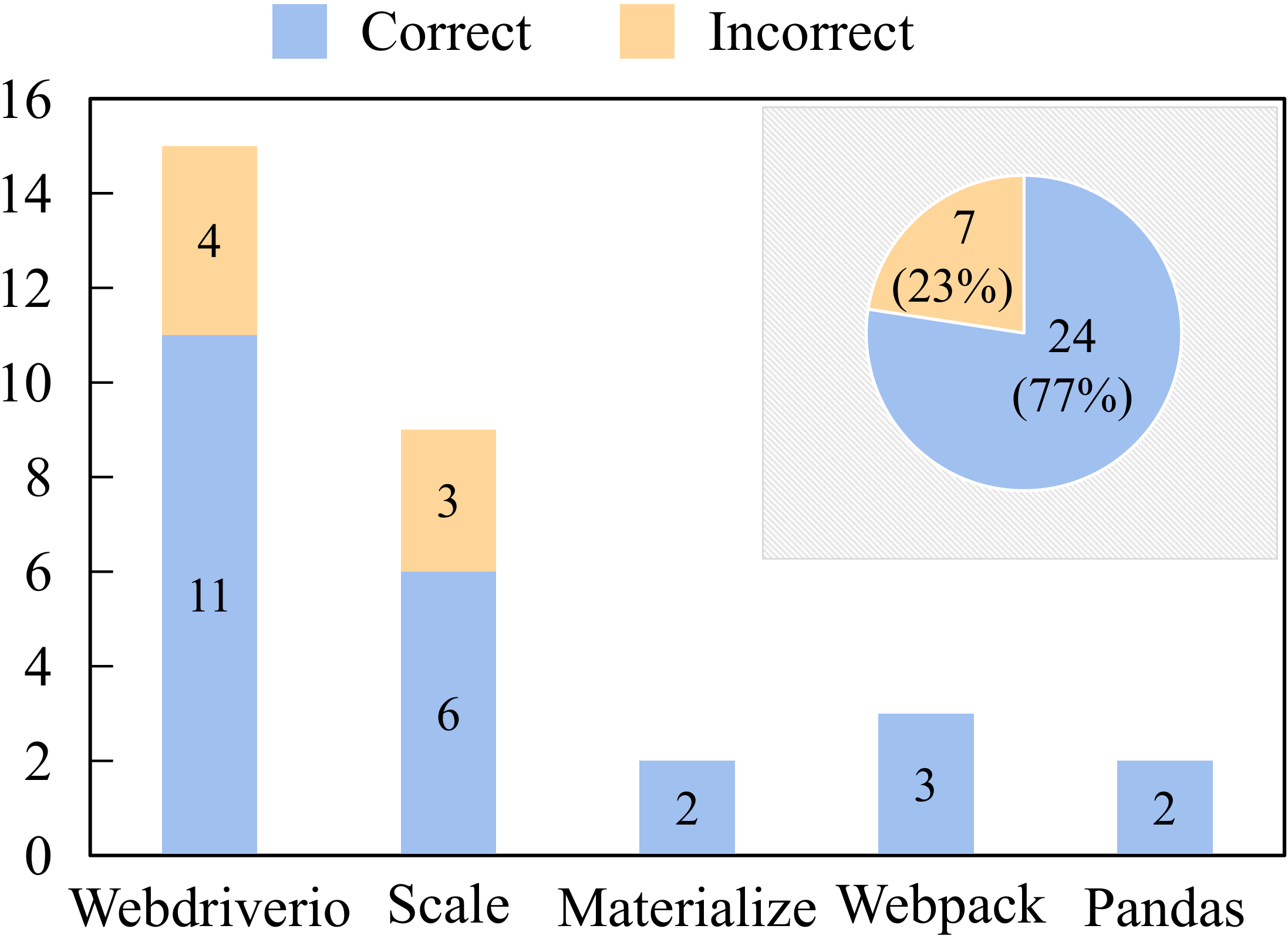}
    
}
\subfigure[Quality and usefulness]{
        \label{fig:Quality and usefulness}
        \includegraphics[width=0.8\columnwidth]{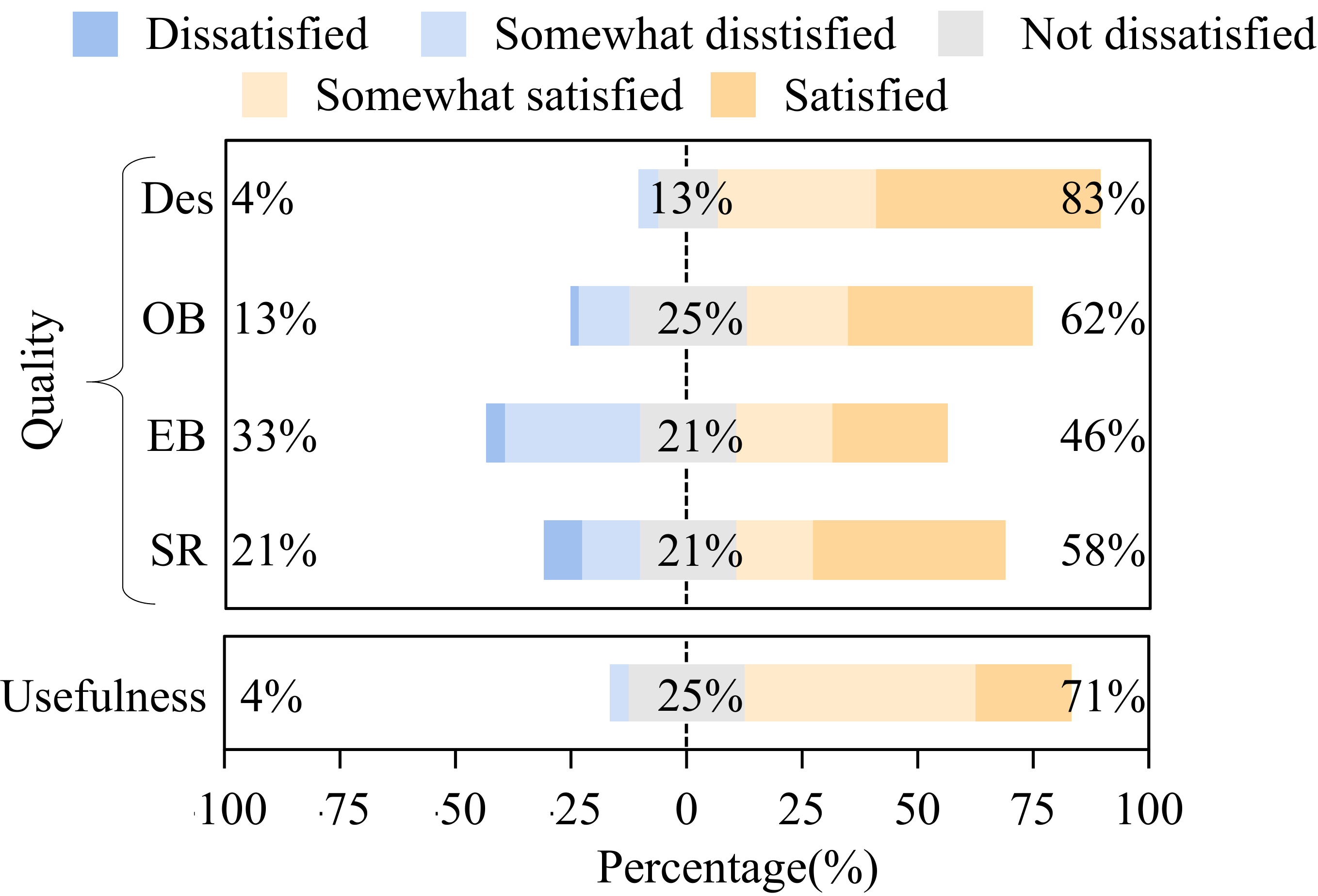}
    
}
\vspace{-0.4cm}
\caption{Results of human evaluation}
\label{fig:Human evaluation}
\vspace{-0.35cm}
\end{figure}

\textbf{Results.} 
{To validate the correctness of bug reports identified by {\tool}, we {ask each participant to} determine whether it is a real bug report {and aggregate group decision} based on the majority vote from the three participants. To validate the quality and usefulness of each identified bug report, we {ask each participant to rate using a scheme from 1-10 and} use the average score of the three evaluations as the final score.}
Fig. \ref{fig:Correctness} shows the bar and pie chart depicting the correctness of {\tool}. Among the 31 bug reports {identified by {\tool}}, 24 (77\%) of them are correct, while 7 (23\%) of them are incorrect.
The correctness is in line with our experiment results (80\% precision of bug report identification). The bar chart shows the correctness distributed among the five communities. The correctness ranges from 63\% to 100\%. The perceived correctness indicates that {\tool} is likely generalized to other open source communities with a relatively good and stable performance.
Fig. \ref{fig:Quality and usefulness} shows an asymmetric stacked bar chart depicting the perceived quality and usefulness of {\tool}'s bug reports, in terms of description, observed behavior, expected behavior, and step to reproduce.
We can see that, the high quality of bug report description is {highly} admitted, 
85\% of the responses agree that the bug report description is satisfactory (i.e., ``somewhat satisfied'' or ``satisfied'').
The high quality of OB, EB, and S2R are also {moderately} admitted (62\%, 46\%, and 58\% on aggregated cases, respectively).
In addition, the usefulness bar chart shows that 
{71\%} of participants agree that {\tool} is useful. 
We will further discuss where does {\tool} perform unsatisfactorily in Sec.\ref{sec:discussion:bad}.

\section{Discussion} 
\label{sec:discussion}
Encouraged by the significant advantages of {\tool} as shown in Sec.6, we believe that our approach could facilitate the bug discovering process and software quality improvement. In this section, we
propose potential usage scenarios as well as improvement opportunities for future work.

\subsection{Potential Usage Scenario}
Software Engineering Bots are widely known as convenient ways for workflow streamlining and productivity improvement  \cite{DBLP:conf/sigsoft/ErlenhovN020,DBLP:conf/kbse/KhananLPJTCRS20, DBLP:journals/ese/AbdellatifBS20}.  
{\tool} can be easily incorporated into a collaborative bot on Gitter, following the basic implementation ideas:
first, the OSS repository owner or core team members who care about the potential bugs could subscribe to their interesting chat rooms via {\tool}; then, {\tool} will monitor the corresponding chat rooms and send 
potential bug reports periodically; and finally, for the bug reports that are confirmed by subscribers, {\tool} could automatically pull them to code repositories such as Github or Gitlab that are well integrated with Gitter. We believe that {\tool} could enhance individual and team productivity as well as improving software quality.


\subsection{Improvement Opportunities}
\label{sec:discussion:bad}
As reported in Sec. 6, 7 out of 31 bug reports are incorrectly labeled by {\tool}. 
{To identify further improvement opportunities for follow-up studies,}
we summarized the following {special cases} {based on examining the human evaluation results} that {necessitates further studies to improve the performance of {\tool}.}

\textbf{(1) Dialogs with a few or no feedback.} We found that 5 out of the 7 incorrect cases are related to insufficient feedback, i.e., three monologues, and the other two with less than five utterances in total. When deciding whether a dialog contains a bug or not, the feedback provided by other developers is important. For example, feedback such as ``it is still not working'' and ``could you please file an issue'' likely indicate the discussing bug should be reported. Therefore, it is difficult for {\tool} to predict dialogs with insufficient feedback. In the future, follow-up research can enrich the bug report classification by adding different confidence levels: \textit{High} and \textit{Normal}. ``High'' refers to the bug reports that the reporter or the discussants have confirmed, and ``Normal'' refers to the bug reports that have the potential.


\textbf{(2) {Dialogs reflecting user misuse/mistake.}} 
We observed that 2/7 incorrect bug reports are actually {associated with installation and version-update due to the users' mistake or negligence.} 
The difference between ``Bugs'' and  {``user misuse/mistake''} is subtle. Both of them might contain negative complaints, error stack traces, and similar keywords such as ``I get errors'', ``not addressed at all'', etc. In the future, follow-up studies are needed to incorporate priori knowledge (e.g., dialogs discussing installation, updating, or building issues are likely not reporting bugs.) to better distinguish the two categories.

\subsection{Threats to Validity}
The first threat is generalizability. {\tool} is only evaluated on {\numProject} open-source projects, which might not be representative of closed-source projects or other open-source projects. The results may be different if the model is applied to other projects. However, our dataset comes from {\numProject} different fields. The variety of projects relatively reduce this threat. 

The second threat may come from the results of automated dialog disentanglement. In this study, we manually inspect and correct the disentanglement results to ensure high-quality inputs for evaluating {\tool}. The average correctness is 79\% in our inspection.
However, for the fully automatic usage of {\tool}, the trade-off option would be directly adopting the automated disentanglement results. {Thus, in real-world application scenarios without manual correction, a slight drop in performance might be observed. 
To alleviate the threat, four state-of-the-art disentanglement models are selected and experimented on live chat data. We adopt the best performing model among the four models, the FF model, to disentangle the live chat. The results of human evaluation study show that {\tool} can achieve 77\% precision without manual correction, and the performance only slightly declined by 3\% compared with {\tool} taking the corrected dialogue as input.} 
Therefore, we believe this can serve as a good foundation for {\tool}'s fully automatic usage. 

The third threat relates to the construct of our approach.
First, we hypothesize that the contents of bug reports likely consist of reporters' utterances, which occasionally results in missing context information. To alleviate the threat, we thoroughly analyzed where our approach performs unsatisfactorily in Sec. 7.2, and planned future work for improvement.
Second, we enlarge our BRI dataset by using a heuristic data augmentation, which may alter the semantics of the original dialog. To alleviate the threat, we employ the utterance mutation from two dimensions (utterance-level and word-level), which has been commonly used in augmenting the datasets for NLP tasks \cite{DBLP:conf/acl/FengGWCVMH21, DBLP:conf/emnlp/WeiZ19}. It could reduce semantic changes of the overall dialogs to a minimum.

The fourth threat relates to the suitability of evaluation metrics. We utilize precision, recall, and F1 to evaluate the performance. We use the dialog labels and utterance labels manually labeled as ground truth when calculating the performance metrics. The threats can be largely relieved as all the instances are reviewed with a concluding discussion session to resolve the disagreement in labels based on majority voting.
There is also a threat related to our human evaluation. We cannot guarantee that each score assigned to every bug report is fair. To mitigate this threat, each bug report is evaluated by 3 human evaluators, and we use the average score of the 3 evaluators as the final score.

\section{Related Work}

\textbf{Identifying Bug Reports.}
Identifying bug reports from user feedback timely and precisely is vital for developers to update their applications. 
Many approaches have been proposed to identify bugs or problems from app reviews \cite{DBLP:conf/icse/0002S20,DBLP:conf/kbse/VuNPN15,DBLP:conf/kbse/VuPNN16,maalej2015bug,DBLP:journals/re/MaalejKNS16,DBLP:conf/icse/GaoZLK18,DBLP:conf/icse/GaoZD0ZLK19,DBLP:journals/tse/ScalabrinoBRPO19}, mailing lists \cite{DBLP:conf/kbse/SorboPVPCG15,DBLP:conf/icse/SorboPVPCG16}, and issue requests \cite{Huang2018Automating,DBLP:conf/icse/AryaWGC19,DBLP:conf/icsm/TerdchanakulHPM17,DBLP:journals/alr/Polpinij21,DBLP:conf/iwpc/PerezJUV21}.
For example, 
Vu et al. \cite{DBLP:conf/kbse/VuNPN15} detected emerging mobile bugs and trends by counting negative keywords based on Google Play. 
Maalej et al. \cite{maalej2015bug,DBLP:journals/re/MaalejKNS16} leveraged natural language processing and sentiment analysis techniques to classify app reviews into bug reports, feature requests, user experiences, and ratings. 
Scalabrino et al. \cite{DBLP:journals/tse/ScalabrinoBRPO19} developed CLAP to classify user reviews into bug reports, feature requests, and non-functional issues based on a random forest classifier.
Di Sorbo et al. \cite{DBLP:conf/kbse/SorboPVPCG15,DBLP:conf/icse/SorboPVPCG16} classified sentences in developer mailing lists into six categories: feature request, opinion asking, problem discovery, solution proposal, information seeking, and information giving. 
Huang et al. \cite{Huang2018Automating} addressed the deficiencies of Di Sorbo et al.'s taxonomy by proposing a convolution neural network (CNN)-based approach. 
Our work differs from existing researches in that we focus on identifying bug reports from collaborative live chats, which pose different challenges as chat messages are interleaved, unstructured, informal, and typically have insufficient labeled data than the previously analyzed documents.

\textbf{Synthesizing Bug Reports.}
Several efforts have been made to synthesize bug reports by utilizing heuristic rules automatically \cite{bettenburg2008extracting,DBLP:conf/esem/DaviesR14,zimmermann2010makes,DBLP:conf/sigsoft/BettenburgJSWPZ08}. 
As heuristic approaches often fail to capture the diverse discourse in bug reports, learning-based approaches have been proposed \cite{DBLP:conf/icsr/ZhaoMYZP19,DBLP:conf/sigsoft/ChaparroBLMMPPN19,DBLP:conf/kbse/ChenXYJCX20,song2020bee}.
Song et al. \cite{song2020bee} proposed a tool that integrates three SVM models to identify the observed behavior, expected behavior, and S2R at the sentence level in bug reports.
Zhao et al. \cite{DBLP:conf/icsr/ZhaoMYZP19} proposed an SVM-based approach that automatically extracts the textual description of steps to reproduce (S2R) from bug reports. 
Chaparro et al. \cite{DBLP:conf/sigsoft/ChaparroBLMMPPN19} proposed a sequence-labeling-based approach that automatically assesses the quality of S2R in bug reports. 
Chen et al. \cite{DBLP:conf/kbse/ChenXYJCX20} proposed a seq2seq-based approach that automatically generates titles regarding the textual bodies written in bug reports.
Most of these methods focus on structuring or synthesizing bug reports from textual descriptions that depicting bugs in a single-party style, while our 
approach targets to automatically structure and synthesize bug reports from multi-party conversations, complementing the existing studies on a novel resource.

\textbf{Knowledge  Extraction from Collaborative Live Chats.}
Recently, more and more work has realized that collaborative live chats play an increasingly significant role in software development, and are a rich and untapped source for valuable information about the software system \cite{chatterjee2019exploratory,DBLP:conf/cscw/LinZSS16,DBLP:conf/msr/ChatterjeeDKP20}. 
Several studies are focusing on extracting knowledge from collaborative live chats. 
Chatterjee et al. \cite{DBLP:conf/icse/ChatterjeeDP21} 
automatically collected opinion-based Q\&A from online developer chats.
Shi et al. \cite{DBLP:conf/icse/ShiXLWL020} proposed an approach to detect feature-request dialogues from developer chat messages via the deep siamese network.  
Qu et al. \cite{InforSeek_User_Intent_Pred} utilized classic machine learning methods to predict user intent with an average F1 of 0.67. 
Rodeghero et al. \cite{DBLP:conf/icse/RodegheroJAM17} presented a technique for automatically extracting information relevant to user stories from recorded conversations.
Chowdhury and Hindle \cite{chowdhury2015mining} filtered out off-topic discussions in programming IRC channels by engaging Stack Overflow discussions. 
The findings of previous work motivate the work presented in this paper. Our study is different from the previous work as we focus on identifying and synthesizing bug reports from massive chat messages that would be important and valuable information for software evolution.
In addition, our work complements the existing studies on knowledge extraction from developer conversations.

\vspace{-0.1cm}
\section{Conclusion}
In this paper, we proposed a novel approach, named {\tool}, which can automatically identify and synthesize bug reports from live chat messages. {\tool} leverages a novel graph neural network to model the graph-structured information of dialog, thereby effectively predicts the bug-report dialogs. {\tool} also adopts {a twice fine-tuned BERT} model by incorporating the transfer learning technique to synthesize complete bug reports. The evaluation results show that our approach significantly outperforms all other baselines in both BRI and BRS tasks.
We also conduct a human evaluation to assess the correctness and quality of the bug reports generated by {\tool}. We apply {\tool} on recent live chats from five new communities and obtain 31 potential bug reports in total. Among the 31 bug reports, 77\% of them are correct. 71\% of human evaluators agree that {\tool} is useful. These results demonstrate the significant potential of applying {\tool} in community-based software development, for promoting bug discovery and quality improvement.

\balance
\section*{Acknowledgments}
We deeply appreciate anonymous reviewers for their constructive and insightful suggestions towards improving this manuscript.
This work is supported by the National Key Research and Development Program of China under Grant No. 2018YFB1403400, the National Science Foundation of China under Grant No. 61802374, 62002348, 62072442, 614220920020 and Youth Innovation Promotion Association Chinese Academy of Sciences.

\bibliographystyle{ACM-Reference-Format}
\bibliography{ref}

\end{document}